\documentclass[journal,twoside,web]{ieeecolor}
\usepackage{tmi}
\usepackage{amsmath,amssymb,amsfonts}
\usepackage{algorithmic}
\usepackage{graphicx}
\usepackage{textcomp}
\usepackage{float}

\usepackage{framed,multirow}
\usepackage{url}
\usepackage{hyperref}
\usepackage{etoolbox}
\DeclareMathOperator*{\argmax}{arg\,max}
\usepackage{mathtools}
\usepackage{filecontents,catchfile}
\usepackage[ruled]{algorithm2e}
\usepackage{subfigure}

\usepackage[style=ieee]{biblatex}
\begin{document}
\title{Semi-weakly-supervised neural network training for medical image registration}
\author{
Yiwen Li,
Yunguan Fu,
Iani J.M.B. Gayo,
Qianye Yang,
Zhe Min,
Shaheer U. Saeed,
Wen Yan,
Yipei Wang,
J. Alison Noble,
Mark Emberton,
Matthew J. Clarkson,
Dean C. Barratt,
Victor A. Prisacariu,
Yipeng Hu
\thanks{Y.Li and V.A. Prisacariu are with Active Vision Laboratory, University of Oxford, Oxford OX1 3PJ, UK(e-mail:yiwen.li@st-annes.ox.ac.uk)}
\thanks{Y.Fu, I.J.M.B.Gayo, Q.Yang, Z.Min, S.U.Saeed, W.Yan, Y.Wang, M.J.Clarkson, D.C.Barratt and Y.Hu are with the UCL centre for Medical Image Computing, Department of Medical Physics and Biomedical Engineering, University College London, London WC1E 6BT, UK and also with the Wellcome/EPSRC Centre for Interventional and Surgical Sciences University College London, London WC1E 6BT, UK}
\thanks{Y.Fu is with InstaDeep Ltd., London W2 1AY, UK}
\thanks{Y.Wen is with Department of Electrical Engineering, City University of Hong Kong, Hong Kong, China}
\thanks{Y.Wang and J.A,Noble are with Institute of Biomedical Engineering, Department of Engineering Science, University of Oxford, Oxford, UK}
\thanks{M.Emberton is with Division of Surgery \& Interventional Science, University College London, London WC1E 6BT, UK}
}
% \author{First A. Author, \IEEEmembership{Fellow, IEEE}, Second B. Author,and Third C. Author, Jr., \IEEEmembership{Member, IEEE}
% % This paragraph of the first footnote will contain the date on which you submitted your paper for review. It will also contain support information, including sponsor and financial support acknowledgment. For example, ``This work was supported in part by the U.S. Department of Commerce under Grant BS123456.''
% \thanks{}
% % The next few paragraphs should contain the authors' current affiliations, including current address and e-mail. For example, F. A. Author is with the National Institute of Standards and Technology, Boulder, CO 80305 USA (e-mail:author@boulder.nist.gov). 
% \thanks{}
% % S. B. Author, Jr., was with Rice University, Houston, TX 77005 USA. He is now with the Department of Physics, Colorado State University, Fort Collins, CO 80523 USA (e-mail: author@lamar.colostate.edu).
% \thanks{}
% % T. C. Author is with the Electrical Engineering Department, University of Colorado, Boulder, CO 80309 USA, on leave from the National Research Institute for Metals, Tsukuba, Japan (e-mail: author@nrim.go.jp).
% \thanks{}}

\maketitle
\begin{abstract}
%%%
%Pairwise image registration is important in a various of clinical applications.
For training registration networks, weak supervision from segmented corresponding regions-of-interest (ROIs) have been proven effective for \textit{a)} supplementing unsupervised methods, and \textit{b)} being used independently in registration tasks in which unsupervised losses are unavailable or ineffective. This correspondence-informing supervision entails cost in annotation that requires significant specialised effort. This paper describes a semi-weakly-supervised registration pipeline that improves the model performance, when only a small corresponding-ROI-labelled dataset is available, by exploiting unlabelled image pairs. We examine two types of augmentation methods by perturbation on network weights and image resampling, such that consistency-based unsupervised losses can be applied on unlabelled data. The novel WarpDDF and RegCut approaches are proposed to allow commutative perturbation between an image pair and the predicted spatial transformation (i.e. respective input and output of registration networks), distinct from existing perturbation methods for classification or segmentation. Experiments using 589 male pelvic MR images, labelled with eight anatomical ROIs, show the improvement in registration performance and the ablated contributions from the individual strategies. Furthermore, this study attempts to construct one of the first computational atlases for pelvic structures, enabled by registering inter-subject MRs, and quantifies the significant differences due to the proposed semi-weak supervision with a discussion on the potential clinical use of example atlas-derived statistics.
%%%%
\end{abstract}

%\linenumbers

%% main text

\newcommand{\codeurl}{\url{https://github.com/kate-sann5100/LowerPelvicReg}}
\newcommand{\dataurl}{\url{https://zenodo.org/record/7013610}}

\section{Introduction}
\subsection{Learning-based medical image registration}
Pairwise image registration is the process of establishing 2D or 3D spatial correspondence between two images, often termed moving and fixed images, such that the estimated spatial-correspondence-representing transformation may be used to align the moving image to the fixed image. 

Such correspondence can be useful in many clinical applications. Inter-subject image registration is the basis of the atlas-based segmentation algorithms and population studies that are important for epidemiology research~\cite{klein2009evaluation}. Intra-subject longitudinal image registration can track suspicious pathological regions for individuals by aligning images acquired from different time points~\cite{yang2020longitudinal}. Registering images from different modalities can combine complementary diagnostic information from different imaging sources to make informed clinical decisions or guide surgery and interventions~\cite{hu2012mr}. Stemmed from computer vision algorithms such as optical flow and pose estimation, proving its usefulness in medical image analysis, registration algorithms have been developed and adopted by many multidisciplinary researches.

The recent development in deep-learning has promoted the investigation towards \textit{learning-based registration} methods~\cite{haskins2020deep,fu2020deep,tustison2019learning}. Fully supervised training of a registration network requires a number of pairs of moving and fixed images, with corresponding ground-truth transformation. Ground-truth transformations, especially deformable transformations, e.g. with higher degrees of freedom than those in rigid or affine transformations, are scarce and prohibitively infeasible to be reliably obtained from many clinical applications. Therefore many methods adopt unsupervised approaches, which rely on the availability of effective and robust image similarity measures.
Improved performance could be achieved by introducing weak supervision from auxiliary information - manual tracing of hippocampus head and body for intra-patient hippocampus registration~\cite{zhu2020neurreg}, lower pelvic landmarks for intra-patient and multimodel registration~\cite{hu2018weakly,hu2018label,zhu2020neurreg} and left and right ventricle cavity and myocardium for intra-patient cardiac registration~\cite{hering2019enhancing}. In cases, these segmentations of corresponding structures may be sufficient or even beneficial without image similarity being used as part of the loss functions - the cardinal form of weakly-supervised registration~\cite{hu2018weakly,hu2018label}. 

\subsection{Semi-weakly-supervised registration}
First, it is important to clarify ``weakly-supervised'' and ``semi-supervised'' approaches for the purpose of this study, where the supervision comes from the available segmentation of corresponding regions of interest (ROIs), between moving and fixed images. As discussed in the original papers~\cite{hu2018weakly,hu2018label,hering2019enhancing,zhu2020neurreg}, these segmented ROIs represent a different type and ``weaker'' labels of correspondence at the ROI level, whilst the objective of registration network training is to learn (and therefore to predict at inference) dense correspondence at voxel level. Semi-supervised learning has also been used to refer to the same class of registration network training methodologies~\cite{cardoso2022monai}, perhaps emphasising that the required voxel-level supervision are only partially available at a subset of all voxel locations (i.e. segmented ROIs). This is particularly judicious when the segmented ROIs represent point landmarks. However, to avoid terminological confusion, we hereinafter refer this type of segmentation-driven methods to as weakly-supervised training, thus the ``semi-weakly-supervised'' (or semi-supervised in the context of this work) is used to indicate the partially available segmentation labels, that is a proper subset of training images (or subjects) with labelled corresponding ROIs. 

Second, it is also useful to point out that many clinical applications lack a valid and robust image intensity-based similarity measure that can be used for the unsupervised loss between two unlabelled images~\cite{hu2013registration}. It is this challenge in many multi-modal registration problems did the weakly-supervised approaches~\cite{hu2018weakly,hu2018label,zhu2020neurreg} intend to address. In these applications, general training approaches (beyond the existing intensity-similarity-based unsupervised losses) for utilising images without segmented corresponding ROIs is desirable. Although considered outside of the scope of this work, the proposed semi-weakly-supervised approaches, which utilise the ``unlabelled image pairs'', could also be useful in combination with these intensity-similarity-based unsupervised losses, which utilise ``unlabelled voxel''. 

In general, these segmentation labels for assisting registration require expert knowledge of radiological anatomy and pathology as well as a significant amount of time. It is therefore valuable to develop methods that require less annotation effort without minimum sacrifice of performance improvement. 
This paper investigates semi-weakly-supervised registration approaches, which reduce cost in labelling by training on the combination of a small labelled dataset and a large unlabelled dataset, while minimising the registration performance loss. 

It is interesting to postulate that, at the time of writing, recent foundation models~\cite{bommasani2021opportunities} having yet been applied in segmentation-supervised registration tasks may be due to the lack of unsupervised (or self-supervised) training methods for such registration. Utilising a large amount of images without labels has been proven effective or even necessary in foundation model development~\cite{azad2023foundational,nvidia,zhou2023foundation}, and may resort to methodologies developed in this work for constructing and/or adapting foundation models for registration tasks.

\subsection{Commutative property for transformation prediction}
Semi-supervised learning based on consistency regularisation has proven its efficacy on image segmentation - another task requires pixel-level prediction. Such consistency is enforced between predictions with and without various perturbations applied to input images, network weights and architectures.
% image perturbation for segmentation
Image perturbation applies augmentations such as affine transformation, CutOut, CutMix and ClassMix to the input image, and penalises the difference between predictions when the augmentation is applied before and after the inference. 
Specifically, given a randomly sampled augmentation $A$, an image $I$ could be augmented to $\tilde{I}=A(I)$. The model $g$ is trained to minimise the distance between the prediction from the augmented image $M^\text{pre}=g(\tilde{I})=g(A(I))$ and the prediction from the original image after applying the same augmentation $M^\text{post}=A(g(I))$. 

In other words, image perturbation for semi-supervised segmentation assumes an intuitive functional commutative property, between the image transformation (for augmentation) $A$ and segmentation prediction $g$, and minimises the difference between $g(A(I))$ and $A(g(I))$.

% different domain
% community rule not apply
% new consistency equation

% why different for registration
However, learning-based registration networks have different types of model input and output, i.e. predicting the corresponding transformation parameters $U=g(I^\text{mov}, I^\text{fix})$, given a pair of moving and fixed images $(I^\text{mov}, I^\text{fix})$. The model input and output are in two very different domains, the ``image-pair domain'' and the ``transformation domain''.
Therefore, the commutative relationship, between the transformation prediction $g$ and existing augmentation methods $A$ used in segmentation (or other task types), does not apply directly.

This motivates our development of a class of commutative perturbation methods, such that the perturbed model output (predicted transformation) is expectedly equivalent to the model output based on a perturbed input (an image pair to be registered), for consistent losses to be calculated between the two transformations.
This paper proposes two cost-efficient image augmentation techniques and the corresponding consistency rule for semi-weakly-supervised registration. Specifically, for each augmentation $A$ applied to the image pairs $(I^\text{mov}, I^\text{fix})$, for generating an augmented image pairs $(\tilde{I}^\text{mov}, \tilde{I}^\text{fix})=A(I^\text{mov}, I^\text{fix})$, a corresponding augmentation $\tilde{A}$ in the transformation domain is proposed, such that consistency could be enforced by penalising the difference between $U^{pre}=g(A(I^\text{mov}, I^\text{fix}))$ and $U^{post}=\tilde{A}(g(I^\text{mov},I^\text{fix}))$.

\subsection{Summary of contributions}
In this work, we report experimental results using an example application of inter-subject registration between male lower pelvic MR images from prostate cancer patients. The clinical significance of this application is discussed with one of the first atlas construction studies in this area (Sec.~\ref{sec:exp_atlas}). 
\begin{itemize}
    \item We first adopted semi-supervision approaches and established a baseline for comparison, in an inter-subject lower-pelvic MR image registration task.
    \item We proposed two novel commutative image perturbation methods and corresponding consistency rules designed specifically for image registration.
    \item We carried out experiments on inter-patient multi-structural lower-pelvic MR registration which proved the efficacy of the proposed method.
    \item We presented extensive ablation studies on the effect of image and weight perturbation, with an attempt to demonstrate the value of semi-weak supervision in building a first-of-its-kind lower-pelvic MRI atlas.
\end{itemize}

\section{Related Work}
\subsection{Medical image registration}
Early algorithms~\cite{christensen19943d,sprengel1996thin,viola1997alignment,thirion1998image,ashburner1999nonlinear,rueckert1999nonrigid,gefen2003elastic,avants2008symmetric,klein2008optimisation} formulate registration as an iterative optimisation to maximise similarity between fixed image and transformed moving image.

Nowadays, most of state-of-the-art methods adopt deep neural networks trained large training sets and allow single-step inference. Due to the scarcity of ground-truth transformation, especially for deformations, most deep-learning approaches adopted unsupervised or weakly-supervised training which maximises the similarity between fixed image and moving image warped by the predicted transformation. Compared with unsupervised methods that solely depend on image similarity metrics~\cite{de2017end,ghosal2017deep,li2017non,stergios2018linear,sheikhjafari2018unsupervised}, weakly-supervised approaches guide registration with auxiliary information like segmentation maps and landmark achieved better and more robust performance.
Hu et al.~\cite{hu2018weakly,hu2018label} improved multimodal registration performance with weak-supervision from full gland segmentations, lower-pelvic landmarks and patient-specific point landmarks.
Hering et al.~\cite{hering2019enhancing} utilised annotations for left and right ventricle cavity and myocardium of a clinical expert for better intra-patient cardiac registration.
Zhu et al.~\cite{zhu2020neurreg} validated the efficacy of guidance from manual tracing of hippocampus and prostate segmentation respectively for intra-patient hippocampus and prostate registration.

\subsection{Semi-supervised learning}\label{sec:background-semi}
Semi-supervised learning improves model performance without increasing labelling expanse by incorporating a large amount of unlabelled data during training. It is now commonly adopted in image segmentation - another task which requires pixel-level prediction.

Early methods utilised GANs either to introduce auxiliary tasks to discriminate real and unreal images generated by GAN~\cite{souly2017semi}, or to train the model to fool a discriminator which differentiates predictions and ground truth~\cite{hung2018adversarial}.
State-of-the-art methods mostly adopt consistency regularisation, which enforces consistency across predictions with various perturbations. 
Weight perturbation methods~\cite{feng2020semi,ke2020guided,chen2021semi,liu2022perturbed} enforced consistency between predictions from multiple networks with the same architecture but different weights.
Image perturbation methods~\cite{french2019semi,kim2020structured,zou2020pseudoseg,liu2022perturbed} augmented input images with techniques and enforced consistency of predictions between augmented images. Commonly adopted spatial augmentation techniques include classical transformations such as rotation, scaling and translation, as well as more strongly augmentation methods such as CutOut, CutMix, ClassMix and so on. While these augmentations could be directly applied to both image and predicted segmentation to form matched pairs for segmentation tasks, processing is required to reflect these transformation on registration outputs (e.g. dense displacement field). This paper proposed DDF-based transformation which could be directly applied to the registration output and RegCut, a strong augmentation designed for registration.

While registration has been utilised to generate pseudo-labels for unlabelled images in some semi-supervised segmentation algorithms~\cite{ito2019semi,xu2019deepatlas,li2022coupling}, from our knowledge, no prior work adopted semi-supervision directly on the segmentation-supervised registration tasks.

% \cite{feng2020semi} teacher-student
% \cite{ke2020guided,chen2021semi} dual student

\section{Method}

\subsection{Task definition}\label{sec:method-task}

%\subsection{Predicting spatial transformation for registration}\label{sec:registration}
Given a pair of moving image $I^\text{mov}\in\mathbb{R}^{W\times H\times D}$ and fixed image $I^\text{fix}\in\mathbb{R}^{W\times H\times D}$, pairwise registration spatially aligns the moving image towards the fixed image. 
$W$, $H$ and $D$ indicating the width, height and depth of the 3D volumetric images~\footnote{The same size is assumed for both images for notional brevity, without losing its generality in the proposed methods or discussion in this study.}, respectively.
The components of the images~\footnote{In this paper, uppercase $I^\text{mov}$ and $I^\text{fix}$ indicate tensor-valued random variables, whilst lowercase $i^\text{mov}_{xyz}$ and $i^\text{fix}_{xyz}$ are scalar-valued tensor components with the subscripts $x$, $y$ and $z$.}, $i^\text{mov}_{xyz}$ and $i^\text{fix}_{xyz}$, are indexed by their voxel coordinates $x=1,...,W$, $y=1,...,H$ and $z=1,...,D$, representing image intensity at individual voxels, respectively.
Registration networks predict a dense displacement field (DDF), $U\in\mathbb{R}^{3\times W\times H\times D}$ with its components $u_{dxyz}$, where $d\in\{1,2,3\}$, and $u_{1xyz}$, $u_{2xyz}$ and $u_{3xyz}$ indicating the $x$-, $y$- and $z$-displacement at voxel location $(x,y,z)$.
A registration network $g(\cdot;\theta)$ with parameters $\theta$, i.e. network weights, aims to predict a spatial transformation, represented by DDF, from the given image pair:
\begin{equation}\label{eq:reg-net}
    U=g(I^\text{mov},I^\text{fix};\theta)
\end{equation}

For weakly-supervised training, a labelled training set $\mathcal{D}_{(\text{lab})}$ is available, from which moving-fixed image pairs, $I^\text{mov}_{(\text{lab})}$ and $I^\text{fix}_{(\text{lab})}\in\mathbb{R}^{W\times H\times D}$, together with a set of $C$-class segmentation masks, $M^\text{mov}_{(c)}$ and $M^\text{fix}_{(c)}$ of the same image size $\mathbb{R}^{W\times H\times D}$, can be sampled, $(I^\text{mov}_{(\text{lab})},I^\text{fix}_{(\text{lab})},\{M^\text{mov}_{(c)}\},\{M^\text{fix}_{(c)}\}) \sim \mathcal{D}_{(\text{lab})}$, where $C$ indicates the number of classes (here, types of anatomical ROIs) thus $c=1,...,C$. 
The components of $M^\text{mov}_{(c)}$ and $M^\text{fix}_{(c)}$ are indexed as $m^\text{mov}_{(c)xyz}$ and $m^\text{fix}_{(c)xyz}$, respectively. Both are assigned with a value of $1$, if voxel $(x,y,z)$ belongs to the ROI class $c$, $0$ otherwise, i.e. $[m^\text{mov}_{(1)xyz},...,m^\text{mov}_{(C)xyz}]^{\top}$ be a voxel-wise one-hot vector.

Semi-weakly-supervised training has access to an additional unlabelled training set $\mathcal{D}_{(\text{unl})}$, such that the samples of unlabelled moving-fixed image pairs, of the same image size, can be drawn, $(I^\text{mov}_{(\text{unl})}, I^\text{fix}_{(\text{unl})}) \sim \mathcal{D}_{(\text{unl})}$ without segmentation masks.
The following sections describe our proposed methods to utilise both $\mathcal{D}_{(\text{lab})}$ and $\mathcal{D}_{(\text{unl})}$ for training the registration network in Eq.~\ref{eq:reg-net}.

\begin{figure*}
    \centering
    \includegraphics[width=0.7\linewidth]{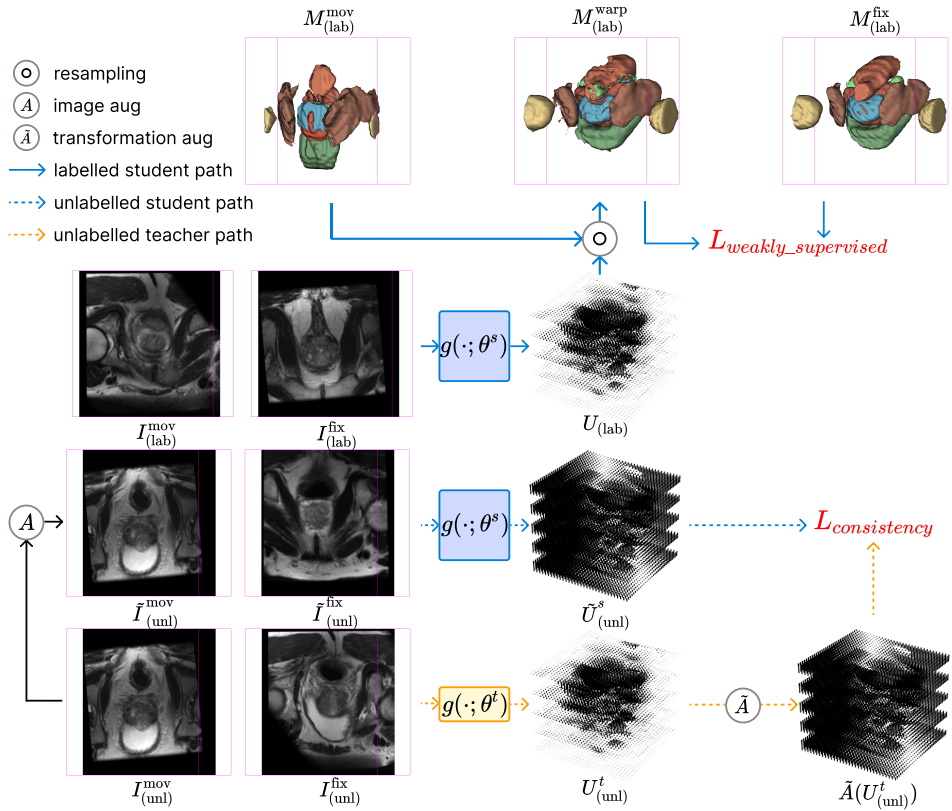}
    \caption{The proposed semi-weakly-supervised registration pipeline with $L_{weakly\_supervised}$ calculated from the labelled pair and $L_\text{cons}$ calculated from the unlabelled pair.}
    \label{fig:pipeline}
\end{figure*}

\subsection{Weak supervision loss on labelled data}\label{sec:method-weak}
A registration network $U_{(\text{lab})}=g(I^\text{mov}_{(\text{lab})}, I^\text{fix}_{(\text{lab})};\theta)$ can be trained with the labelled set $D_{(\text{lab})}$, by minimising a loss function $L_\text{weak-sup}$ between the $U_{(\text{lab})}$-warped moving segmentation masks $M^\text{warp}_{(c)}$ and the fixed masks $M^\text{fix}_{(c))}$, where
\begin{equation}
    M^\text{warp}_{(c)} = M^\text{mov}_{(c)} \circ U_{(\text{lab})}
\end{equation}
where, the image resampling operation is denoted by ``$\circ$'', resampling the left image-sized mask (or image) $M^\text{mov}_{(c)}$ using the right DDF $U_{(\text{lab})}$, such that the indexed values $m^\text{warp}_{(c)xyz}$ at its $(x, y, z)$ resampled voxel locations are spatially-interpolated from the $(x+u_{1xyz}, y+u_{2xyz}, z+u_{3xyz})$ locations from the original image $m^\text{mov}_{(c)xyz}$ coordinates. In practice, the inverse fixed-to-moving DDF is implemented, for resampling at regular fixed image grid locations avoiding numerical inverting.
Substituting the predicted DDF $U_{(\text{lab})}$, we have:
\begin{equation}
    L_\text{weak-sup} = \mathbb{E}_{c}[ L_\text{dice}(M^\text{mov}_{(c)} \circ g(I^\text{mov}_{(\text{lab})}, I^\text{fix}_{(\text{lab})};\theta), M^\text{fix}_{(c)}) ]
\end{equation}
where $\mathbb{E}_{c}(\cdot)$ denotes statistical expectation over $C$ classes. In this work, the Dice loss is used:

\begin{equation}\label{eq:weak-loss}
    \begin{aligned}
        L_\text{dice}&(M^\text{warp}_{(c)},M^\text{mov}_{(c)}) \\
            &= -\frac{2 \cdot\sum_{x,y,z} m^\text{warp}_{(c)xyz} m^\text{mov}_{(c)xyz}}{\sum_{x,y,z}{m^\text{warp}_{(c)xyz}} + \sum_{x,y,z}{m^\text{mov}_{(c)xyz}}}
        %f_\text{dice}(M, \Tilde{M})=\frac{1}{C}\sum_{\lambda} \frac{2\sum_{x,y,z}M_{\lambda xyz}*\Tilde{M}_{\lambda xyz}}{\sum_{x,y,z}M_{\lambda xyz}+\Tilde{M}_{\lambda xyz}}
    \end{aligned}
\end{equation}
 
In experiments described in Sec.~\ref{sec:experiments}, the described weak supervision loss is tested independently without using unlabelled training set as a baseline. It is also used in conjunction of the consistency losses introduced in the following Secs.~\ref{sec:weight_perturbation} and~\ref{sec:image_perturbation} for the proposed semi-weakly supervision.

\subsection{Consistency following weight perturbation}\label{sec:weight_perturbation}
For unlabelled training set $\mathcal{D}_{(\text{unl})}$, network weight perturbation is introduced by adopting the student-teacher paradigm, which includes a student model and a teacher model with the same network architecture but separate weights, as illustrated in Fig.~\ref{fig:pipeline}. During training, the weights of teacher are a temporal ensemble from the weights of student. Denote the student and teacher weights as $\theta$ and $\theta_t$, respectively. The teacher weights are updated through an exponential moving average (EMA) scheme based on the current student weights:
\begin{equation}
    \theta^t = \gamma\times\theta^t+(1-\gamma)\times\theta
\end{equation}
where $\gamma\in(0,1)$ is a hyperparameter controlling the transfer weight between epochs.

Given an pair of unlabelled image $(I^\text{mov}_{(\text{unl})}, I^\text{fix}_{(\text{unl})})$, the student and teacher model respectively predicts $U_{(\text{unl})}^s = g(I^\text{mov}_{(\text{unl})}, I^\text{fix}_{(\text{unl})};\theta)$ and $U_{(\text{unl})}^t = g(I^\text{mov}_{(\text{unl})}, I^\text{fix}_{(\text{unl})};\theta^t)$. Weight perturbation enforces consistency between the two DDFs by a consistency loss function $L_\text{cons}$:
\begin{equation}\label{eq:consistency-loss}
    L_\text{cons} = f_\text{mse}(U_{(\text{unl})}^t, U_{(\text{unl})}^s)
\end{equation}
where $f_\text{mse}$ is a mean-square-error loss defined using the same tensor component notation as in Sec.~\ref{sec:method-task}:
\begin{equation}
    f_\text{mse}(U_{(\text{unl})}^t, U_{(\text{unl})}^s) = \mathbb{E}_{d,x,y,z}[(u^t_{(\text{unl})dxyz} - u^s_{(\text{unl})dxyz})^2]
\end{equation}
The student is trained through backpropagation to minimise the weak supervision loss (Eq.~\ref{eq:weak-loss}) calculated on the labelled pairs and the consistency loss (Eq.~\ref{eq:consistency-loss}) calculated on the unlabelled pairs, sampled from $\mathcal{D}_{(\text{lab})}$ and $\mathcal{D}_{(\text{unl})}$, respectively:
\begin{equation}
    L = L_\text{weak-sup} + \alpha L_\text{cons}
\end{equation}
where $\alpha$ is the consistency loss coefficient. This loss function is considered task-agnostic, adapted from previous semi-supervised classification or segmentation (Sec.~\ref{sec:background-semi}).

\subsection{Image perturbation for registration}\label{sec:image_perturbation}

Image perturbation generates an augmented image pair $(\tilde{I}^\text{mov}_{(\text{unl})}, \tilde{I}^\text{fix}_{(\text{unl})}) = A(I^\text{mov}_{(\text{unl})}, I^\text{fix}_{(\text{unl})})$ from the pair of unlabelled image $(I^\text{mov}_{(\text{unl})}, I^\text{fix}_{(\text{unl})})$. 
%and enforces consistency between predicted DDFs, $U_{(\text{unl})}$ and $\hat{U}_{(\text{unl})}$, respectively for the original and augmented pair. 
During training, teacher model predicts a DDF for the unaugmented pair $U^t_{(\text{unl})}=g(I^\text{mov}_{(\text{unl})}, I^\text{fix}_{(\text{unl})};\theta^t)$ while student model predicts the DDF for the augmented pair $\tilde{U}^s_{(\text{unl})}=g(\tilde{I}^\text{mov}_{(\text{unl})}, \tilde{I}^\text{fix}_{(\text{unl})};\theta)$. Whilst $A$ is an augmentation function on the input image pairs, $\tilde{A}$ is an augmentation function applied on the output DDFs $\tilde{A}(U^t_{(\text{unl})})$, such that the two DDFs $\tilde{U}^s_{(\text{unl})}$ and $\tilde{A}(U^t_{(\text{unl})})$ can be compared and the consistency loss may be enforced between the two:
\begin{equation}
    L_\text{cons} = f_\text{mse}(\tilde{A}(U^t_{(\text{unl})}), \tilde{U}^s_{(\text{unl})})
\end{equation}

In the following subsections, two types of (input) image-domain augmentation $A$ are described, each with its corresponding $\tilde{A}$ augmentation in the (output) transformation-domain, for the proposed semi-weakly supervised training.

\subsubsection{WarpDDF}\label{sec:method_image_warpddf}
\begin{figure}
    \centering
    \includegraphics[width=\linewidth]{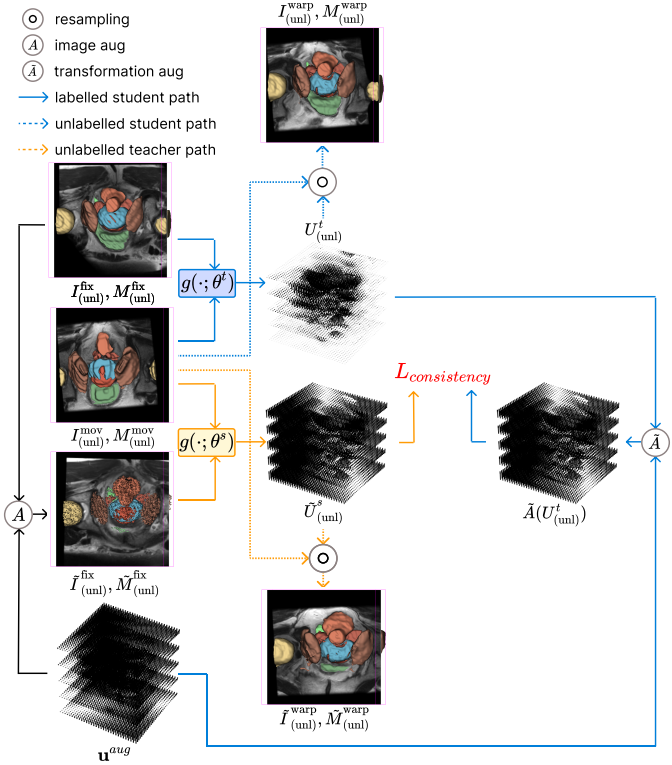}
    \caption{DDF-based augmentation transforms the fixed image $I^\text{fix}_{(\text{unl})}$ by an augmentation dense displacement field $\mathbb{u}^\text{aug}$ and seek consistency between DDFs predicted from the original and augmented pairs. Note segmentation masks are not available during training, they are included here for illustration purpose.}
    \label{fig:DDF_transform}
\end{figure}

Consider a randomly sampled spatial transformation, represented by a DDF $U^\text{aug}\in \mathbb{R}^{3\times W\times H\times D}$, an image-domain augmentation $A$ transforms the fixed image $I^\text{fix}_{(\text{unl})}$ using $U^\text{aug}$, while the moving image $I^\text{mov}_{(\text{unl})}$ remains unchanged, that is:
\begin{equation}
    \begin{aligned}
        \tilde{I}^\text{mov}_{(\text{unl})} &= I^\text{mov}_{(\text{unl})}\\
        \tilde{I}^\text{fix}_{(\text{unl})} &= I^\text{fix}_{(\text{unl})} \circ U^\text{aug}
    \end{aligned}
\end{equation}

Whilst, the corresponding transformation-domain augmentation $\tilde{A}$ becomes:
\begin{equation}\label{eq:aug_ddf}
    \tilde{A}(U^t_{(\text{unl})}) = U^\text{aug} + U^t_{(\text{unl})} \circ U^\text{aug}
\end{equation}
where the $U^t_{(\text{unl})} \circ U^\text{aug}$ operation generalizes the image resampling (described in Sec.~\ref{sec:method-weak}) to the left higher-order-tensor-represented DDF $U^t_{(\text{unl})}$, by independently interpolating each $x$, $y$ and $z$ displacement components (i.e. when $d=1,2,3$, respectively) of $U^t_{(\text{unl})}$, using the right DDF $U^\text{aug}$. 

We provide the following derivation to show that the two DDFs, $\tilde{U}^s_{(\text{unl})}$ and $\tilde{A}(U^t_{(\text{unl})})$, are expected to be consistent. 
\begin{equation}\label{eq:aug_ddf_proof}
    \begin{split}
        \tilde{I}^\text{mov}_{(\text{unl})} \circ \tilde{A}(U^t_{(\text{unl})})
        &=\tilde{I}^\text{mov}_{(\text{unl})} \circ (U^\text{aug} + U^t_{(\text{unl})} \circ U^\text{aug})\\
        &=I^\text{mov}_{(\text{unl})} \circ U^t_{(\text{unl})} \circ U^\text{aug}\\
        &=I^\text{warp}_{(\text{unl})} \circ U^\text{aug}
    \end{split}
\end{equation}

A key insight is that Eq.~\ref{eq:aug_ddf} represents a so-called ``transformation composing'' between $U^t_{(\text{unl})}$ and $U^\text{aug}$ (further details in Appendix~\ref{sec:app_ddf_compose}), which is decomposed and applied to the unchanged moving image $I^\text{mov}_{(\text{unl})}$ in Eq.~\ref{eq:aug_ddf_proof}. This results in applying $U^\text{aug}$ on the warped moving image $I^\text{warp}_{(\text{unl})}$, which is expected to be aligned with the augmented fixed image $\tilde{I}^\text{fix}_{(\text{unl})}=I^\text{fix}_{(\text{unl})} \circ U^\text{aug}$.

The proposed WarpDDF data augmentation pair $A$ and $\tilde{A}$ is general for representing different types of spatial transformation, including rigid, affine and nonrigid variants. This study adopts random rotation, translation and scaling as an example of the proposed WarpDDF in the experiments reported in Sec.~\ref{sec:experiments}, as plausible higher-order nonrigid transformation for inter-subject registration remains an open research question. 

\subsubsection{RegCut}\label{sec:method_image_regcut}
\begin{figure}
    \centering
    \includegraphics[width=\linewidth]{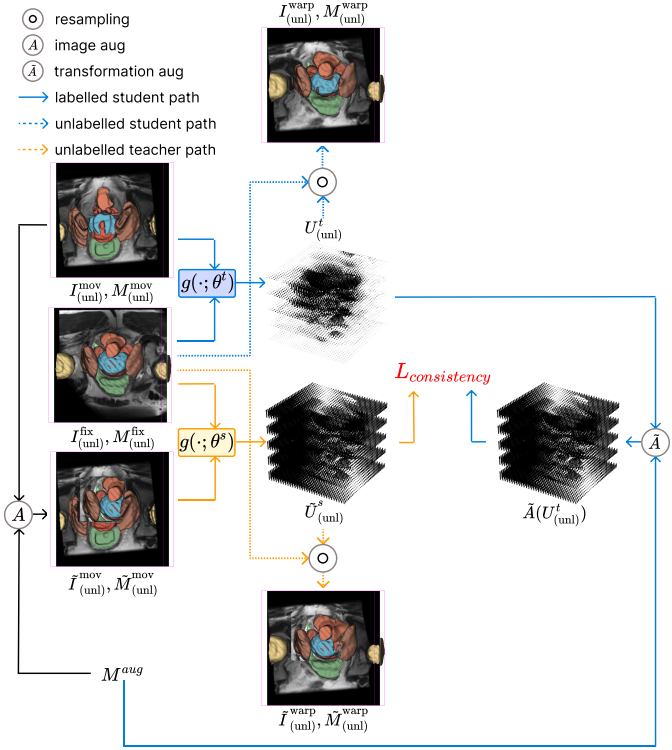}
    \caption{RegCut mix unlabelled moving image with unlabelled fixed image following a randomly sampled mask and seek consistency between DDFs predicted from the original and augmented pairs. Note segmentation masks are not available during training, they are included here for illustration purpose.}
    \label{fig:regcut}
\end{figure}
RegCut augments the training data by mixing unlabelled moving images with unlabelled fixed images, using a random binary mask $M^\text{aug} \in \{1,0\}^{W\times H\times D}$, indexed with $m^\text{aug}_{x,y,z}$. In each training iteration, a rectangular cuboid $B = (r_x,r_y,r_z,r_w,r_h,r_d)$ is sampled, inside which the mask voxels $m^\text{aug}_{xyz}$ are labelled as $1$, i.e. if $x\in[r_x,r_x+r_w)$, $y\in[r_y,r_y+r_h)$ and $x\in[r_z,r_z+r_d)$, $0$ otherwise.

The mask $M^\text{aug}$ is used to generate an augmented moving image $\tilde{I}^\text{mov}_{(\text{unl})}$, by replacing intensity values with those from the fixed image $I^\text{fix}_{(\text{unl})}$ at voxel locations defined by $M^\text{aug}$, while, in this case, the fixed image remains unchanged:
\begin{equation}\label{eq:regcut_image}
    \begin{aligned}
        \tilde{I}^\text{mov}_{(\text{unl})} &= M^\text{aug} \odot I^\text{fix}_{(\text{unl})} + (\mathbf{1}-M^\text{aug}) \odot I^\text{mov}_{(\text{unl})}\\
        \tilde{I}^\text{fix}_{(\text{unl})} &=I^\text{fix}_{(\text{unl})}
    \end{aligned}
\end{equation}
where $\odot$ is component (element)-wise multiplication and $\mathbf{1}$ is a constant-valued tensor of the same image size.

The corresponding transformation-domain augmentation $\tilde{A}(U^t_{(\text{unl})})$ is defined as follows:
\begin{equation}\label{eq:regcut_ddf}
    \tilde{A}(U^t_{(\text{unl})})=(\mathbf{1}-M^\text{aug}) \odot U^t_{(\text{unl})}
\end{equation}
To show the consistency expected between the augmentation pair $A$ and $\tilde{A}$ (defined in Eqs.~\ref{eq:regcut_image} and~\ref{eq:regcut_ddf}) in the proposed RegCut, we compare the augmented moving image warped by the augmented transformation $\tilde{I}^\text{mov}_{(\text{unl})} \circ \tilde{A}(U^t_{(\text{unl})})$ and the augmented fixed image $\tilde{I}^\text{fix}_{(\text{unl})}=I^\text{fix}_{(\text{unl})}$:
\begin{equation}\label{eq:regcut_ddf_proof}
    \begin{split}
        &\tilde{I}^\text{mov}_{(\text{unl})} \circ \tilde{A}(U^t_{(\text{unl})})\\
        =&(M^\text{aug}\odot I^\text{fix}_{(\text{unl})} + (\mathbf{1}-M^\text{aug})\odot I^\text{mov}_{(\text{unl})}) \circ (\mathbf{1}-M^\text{aug}) \odot U^t_{(\text{unl})}\\
        =&M^\text{aug} \odot I^\text{fix}_{(\text{unl})}+(\mathbf{1}-M^\text{aug}) \odot (I^\text{mov}_{(\text{unl})} \circ U^t_{(\text{unl})})\\
        =&M^\text{aug} \odot I^\text{fix}_{(\text{unl})}+(\mathbf{1}-M^\text{aug}) \odot I^\text{warp}_{(\text{unl})}
    \end{split}
\end{equation}
where $I^\text{warp}_{(\text{unl})}=I^\text{mov}_{(\text{unl})} \circ U^t_{(\text{unl})}$ indicates the warped moving image. Eq.~\ref{eq:regcut_ddf_proof} shows that both regions inside and outside of the mask, $M^\text{aug}$ and $\mathbf{1}-M^\text{aug}$, are expected to be consistent between the image- and transformation-domain augmentation.

\subsubsection{Combining WarpDDF and RegCut}\label{sec:method_image_combined}
By design, the two proposed WarpDDF and RegCut augmentation $A$ can readily be combined together in the input image domain:
\begin{equation}
    \begin{aligned}
        \tilde{I}^\text{fix}_{(\text{unl})} &= T(I^\text{fix}_{(\text{unl})},u^\text{aug})\\
        \tilde{I}^\text{mov}_{(\text{unl})} &= M^\text{aug}\odot \tilde{I}^\text{fix}_{(\text{unl})} + (\mathbf{1}-M^\text{aug})\odot I^\text{mov}_{(\text{unl})}
    \end{aligned}
\end{equation}
Using the similar derivation in Eqs.~\ref{eq:aug_ddf_proof} and~\ref{eq:regcut_ddf_proof}, the corresponding output transformation-domain augmentation thus is:
\begin{equation}
    \tilde{A}(U^t_{(\text{unl})})=(\mathbf{1}-M) \odot (U^\text{aug} + U^t_{(\text{unl})} \circ U^\text{aug})
\end{equation}

\section{Experiments}\label{sec:experiments}

\newcommand{\Dicebyclass}{\begin{tabular}{c|c|ccccccccc}%
\hline%
labelled&\multirow{3}{*}{method}&\multicolumn{9}{c}{Dice}\\%
ratio&&\multirow{2}{*}{Bladder}&\multirow{2}{*}{Bone}&Obturator&Peripheral&Central&\multirow{2}{*}{Rectum}&Seminal&Neurovascular&\multirow{2}{*}{Mean}\\%
(\%)&&&&Internus&Zone&Gland&&Vesicle&Bundle&\\%
\hline%
\multirow{1}{*}{0.0}&NifityReg&9.87$\pm$4.53&7.61$\pm$3.65&17.53$\pm$4.03&6.66$\pm$3.39&7.58$\pm$3.98&43.26$\pm$5.27&3.51$\pm$3.21&3.74$\pm$2.87&12.47\\%
\hline%
\multirow{5}{*}{10.0}&sup only&73.90$\pm$4.32&83.20$\pm$3.28&74.95$\pm$2.33&57.59$\pm$4.07&75.84$\pm$3.58&83.07$\pm$2.88&44.70$\pm$4.31&37.82$\pm$3.88&66.38\\%
&NoAug&74.36$\pm$4.64&87.46$\pm$3.15&79.44$\pm$2.11&63.20$\pm$3.95&78.87$\pm$3.72&84.57$\pm$3.17&49.68$\pm$4.74&43.68$\pm$4.29&70.16\\%
&WarpDDF&78.50$\pm$4.30&89.49$\pm$3.03&80.73$\pm$2.16&65.86$\pm$4.03&81.05$\pm$3.26&86.86$\pm$2.62&53.71$\pm$4.73&47.16$\pm$4.31&72.92\\%
&RegCut&77.02$\pm$4.68&88.43$\pm$3.53&80.21$\pm$2.17&65.65$\pm$3.89&80.04$\pm$3.52&86.07$\pm$2.75&55.60$\pm$4.67&45.56$\pm$4.22&72.32\\%
&WarpDDF+RegCut&78.04$\pm$4.57&89.56$\pm$3.26&81.23$\pm$2.11&66.05$\pm$4.00&79.58$\pm$3.50&87.50$\pm$2.79&55.10$\pm$4.76&46.99$\pm$4.44&73.01\\%
\hline%
\multirow{2}{*}{20.0}&sup only&77.20$\pm$4.42&83.71$\pm$3.28&76.19$\pm$2.16&61.65$\pm$3.73&77.34$\pm$3.53&83.80$\pm$2.94&50.97$\pm$4.19&43.26$\pm$3.86&69.26\\%
&WarpDDF+RegCut&75.68$\pm$4.47&86.31$\pm$3.15&78.63$\pm$2.25&65.98$\pm$3.49&78.89$\pm$3.77&85.77$\pm$2.62&54.71$\pm$4.69&49.44$\pm$3.70&71.93\\%
\hline%
\multirow{2}{*}{50.0}&sup only&85.15$\pm$3.06&83.89$\pm$3.42&76.07$\pm$2.05&68.84$\pm$2.76&83.38$\pm$2.67&86.88$\pm$2.20&58.69$\pm$3.83&48.98$\pm$3.48&73.99\\%
&WarpDDF+RegCut&82.55$\pm$3.88&85.14$\pm$3.36&77.87$\pm$1.96&69.45$\pm$2.79&84.49$\pm$2.41&87.12$\pm$2.30&59.73$\pm$3.99&50.68$\pm$3.78&74.63\\%
\hline%
\multirow{1}{*}{100.0}&sup only&83.32$\pm$3.81&85.96$\pm$3.38&79.56$\pm$1.94&69.65$\pm$2.90&84.17$\pm$2.52&87.47$\pm$2.27&60.68$\pm$4.20&51.49$\pm$3.81&75.29\\%
\hline%
\end{tabular}%
}
\begin{table*}[t!]
    \centering
    \begin{tabular}{c|c|ccccccccc}%
\hline%
labelled&\multirow{3}{*}{method}&\multicolumn{9}{c}{Dice}\\%
ratio&&\multirow{2}{*}{Bladder}&\multirow{2}{*}{Bone}&Obturator&Peripheral&Central&\multirow{2}{*}{Rectum}&Seminal&Neurovascular&\multirow{2}{*}{Mean}\\%
(\%)&&&&Internus&Zone&Gland&&Vesicle&Bundle&\\%
\hline%
\multirow{1}{*}{0.0}&NifityReg&9.87$\pm$4.53&7.61$\pm$3.65&17.53$\pm$4.03&6.66$\pm$3.39&7.58$\pm$3.98&43.26$\pm$5.27&3.51$\pm$3.21&3.74$\pm$2.87&12.47\\%
\hline%
\multirow{5}{*}{10.0}&sup only&73.90$\pm$4.32&83.20$\pm$3.28&74.95$\pm$2.33&57.59$\pm$4.07&75.84$\pm$3.58&83.07$\pm$2.88&44.70$\pm$4.31&37.82$\pm$3.88&66.38\\%
&NoAug&74.36$\pm$4.64&87.46$\pm$3.15&79.44$\pm$2.11&63.20$\pm$3.95&78.87$\pm$3.72&84.57$\pm$3.17&49.68$\pm$4.74&43.68$\pm$4.29&70.16\\%
&WarpDDF&78.50$\pm$4.30&89.49$\pm$3.03&80.73$\pm$2.16&65.86$\pm$4.03&81.05$\pm$3.26&86.86$\pm$2.62&53.71$\pm$4.73&47.16$\pm$4.31&72.92\\%
&RegCut&77.02$\pm$4.68&88.43$\pm$3.53&80.21$\pm$2.17&65.65$\pm$3.89&80.04$\pm$3.52&86.07$\pm$2.75&55.60$\pm$4.67&45.56$\pm$4.22&72.32\\%
&WarpDDF+RegCut&78.04$\pm$4.57&89.56$\pm$3.26&81.23$\pm$2.11&66.05$\pm$4.00&79.58$\pm$3.50&87.50$\pm$2.79&55.10$\pm$4.76&46.99$\pm$4.44&73.01\\%
\hline%
\multirow{2}{*}{20.0}&sup only&77.20$\pm$4.42&83.71$\pm$3.28&76.19$\pm$2.16&61.65$\pm$3.73&77.34$\pm$3.53&83.80$\pm$2.94&50.97$\pm$4.19&43.26$\pm$3.86&69.26\\%
&WarpDDF+RegCut&75.68$\pm$4.47&86.31$\pm$3.15&78.63$\pm$2.25&65.98$\pm$3.49&78.89$\pm$3.77&85.77$\pm$2.62&54.71$\pm$4.69&49.44$\pm$3.70&71.93\\%
\hline%
\multirow{2}{*}{50.0}&sup only&85.15$\pm$3.06&83.89$\pm$3.42&76.07$\pm$2.05&68.84$\pm$2.76&83.38$\pm$2.67&86.88$\pm$2.20&58.69$\pm$3.83&48.98$\pm$3.48&73.99\\%
&WarpDDF+RegCut&82.55$\pm$3.88&85.14$\pm$3.36&77.87$\pm$1.96&69.45$\pm$2.79&84.49$\pm$2.41&87.12$\pm$2.30&59.73$\pm$3.99&50.68$\pm$3.78&74.63\\%
\hline%
\multirow{1}{*}{100.0}&sup only&83.32$\pm$3.81&85.96$\pm$3.38&79.56$\pm$1.94&69.65$\pm$2.90&84.17$\pm$2.52&87.47$\pm$2.27&60.68$\pm$4.20&51.49$\pm$3.81&75.29\\%
\hline%
\end{tabular}%
    \label{tab:Dice_by_class}
    \caption{Dice(\%) of the proposed algorithms and its variants at when labels are available for 10\%, 20\%, 50\% and all training images.}
\end{table*}

\newcommand{\HDbyclass}{\begin{tabular}{c|c|ccccccccc}%
\hline%
labelled&\multirow{3}{*}{method}&\multicolumn{9}{c}{95\%HD}\\%
ratio&&\multirow{2}{*}{Bladder}&\multirow{2}{*}{Bone}&Obturator&Peripheral&Central&\multirow{2}{*}{Rectum}&Seminal&Neurovascular&\multirow{2}{*}{Mean}\\%
(\%)&&&&Internus&Zone&Gland&&Vesicle&Bundle&\\%
\hline%
\multirow{1}{*}{0.0}&NifityReg&9.48$\pm$4.27&42.77$\pm$6.52&29.95$\pm$3.46&30.95$\pm$4.34&17.84$\pm$4.45&24.79$\pm$3.28&14.82$\pm$4.42&37.44$\pm$3.95&26.01\\%
\hline%
\multirow{5}{*}{10.0}&sup only&24.92$\pm$3.06&24.81$\pm$3.15&20.63$\pm$2.72&19.98$\pm$2.87&21.84$\pm$3.05&23.96$\pm$2.79&23.79$\pm$2.89&24.18$\pm$2.89&23.01\\%
&NoAug&23.68$\pm$2.96&23.62$\pm$3.10&19.55$\pm$2.66&19.42$\pm$2.93&20.92$\pm$2.93&25.20$\pm$2.78&21.70$\pm$2.86&23.72$\pm$2.99&22.23\\%
&WarpDDF&23.62$\pm$2.96&24.22$\pm$3.31&19.74$\pm$2.61&19.03$\pm$2.80&21.21$\pm$2.95&22.84$\pm$2.67&21.99$\pm$2.95&22.42$\pm$2.90&21.89\\%
&RegCut&27.33$\pm$3.33&25.64$\pm$3.36&20.55$\pm$2.69&18.53$\pm$2.84&20.72$\pm$3.00&24.03$\pm$2.74&22.56$\pm$3.05&23.00$\pm$2.93&22.79\\%
&WarpDDF+RegCut&23.63$\pm$2.98&23.96$\pm$3.26&20.03$\pm$2.64&18.54$\pm$2.79&20.52$\pm$2.90&22.56$\pm$2.66&22.29$\pm$3.00&21.52$\pm$2.96&21.63\\%
\hline%
\multirow{2}{*}{20.0}&sup only&23.79$\pm$3.02&25.35$\pm$3.24&20.48$\pm$2.71&19.87$\pm$2.84&22.82$\pm$3.06&24.32$\pm$2.60&23.22$\pm$2.93&23.67$\pm$2.79&22.94\\%
&WarpDDF+RegCut&24.10$\pm$3.02&24.35$\pm$3.11&19.69$\pm$2.63&19.60$\pm$2.91&21.89$\pm$3.11&23.96$\pm$2.64&22.34$\pm$2.99&23.23$\pm$2.76&22.40\\%
\hline%
\multirow{2}{*}{50.0}&sup only&23.76$\pm$3.12&25.32$\pm$3.22&20.86$\pm$2.70&19.81$\pm$2.79&21.53$\pm$2.99&22.85$\pm$2.74&22.49$\pm$2.95&22.47$\pm$2.72&22.39\\%
&WarpDDF+RegCut&23.36$\pm$3.09&25.02$\pm$3.23&20.31$\pm$2.65&19.40$\pm$2.82&21.43$\pm$2.99&23.91$\pm$2.86&22.75$\pm$3.01&21.71$\pm$2.83&22.24\\%
\hline%
\multirow{1}{*}{100.0}&sup only&23.41$\pm$3.07&25.30$\pm$3.25&20.21$\pm$2.65&19.21$\pm$2.78&21.36$\pm$3.02&23.50$\pm$2.85&22.00$\pm$2.93&21.37$\pm$2.78&22.05\\%
\hline%
\end{tabular}%
}
\begin{table*}[t!]
    \centering
    \begin{tabular}{c|c|ccccccccc}%
\hline%
labelled&\multirow{3}{*}{method}&\multicolumn{9}{c}{95\%HD}\\%
ratio&&\multirow{2}{*}{Bladder}&\multirow{2}{*}{Bone}&Obturator&Peripheral&Central&\multirow{2}{*}{Rectum}&Seminal&Neurovascular&\multirow{2}{*}{Mean}\\%
(\%)&&&&Internus&Zone&Gland&&Vesicle&Bundle&\\%
\hline%
\multirow{1}{*}{0.0}&NifityReg&9.48$\pm$4.27&42.77$\pm$6.52&29.95$\pm$3.46&30.95$\pm$4.34&17.84$\pm$4.45&24.79$\pm$3.28&14.82$\pm$4.42&37.44$\pm$3.95&26.01\\%
\hline%
\multirow{5}{*}{10.0}&sup only&24.92$\pm$3.06&24.81$\pm$3.15&20.63$\pm$2.72&19.98$\pm$2.87&21.84$\pm$3.05&23.96$\pm$2.79&23.79$\pm$2.89&24.18$\pm$2.89&23.01\\%
&NoAug&23.68$\pm$2.96&23.62$\pm$3.10&19.55$\pm$2.66&19.42$\pm$2.93&20.92$\pm$2.93&25.20$\pm$2.78&21.70$\pm$2.86&23.72$\pm$2.99&22.23\\%
&WarpDDF&23.62$\pm$2.96&24.22$\pm$3.31&19.74$\pm$2.61&19.03$\pm$2.80&21.21$\pm$2.95&22.84$\pm$2.67&21.99$\pm$2.95&22.42$\pm$2.90&21.89\\%
&RegCut&27.33$\pm$3.33&25.64$\pm$3.36&20.55$\pm$2.69&18.53$\pm$2.84&20.72$\pm$3.00&24.03$\pm$2.74&22.56$\pm$3.05&23.00$\pm$2.93&22.79\\%
&WarpDDF+RegCut&23.63$\pm$2.98&23.96$\pm$3.26&20.03$\pm$2.64&18.54$\pm$2.79&20.52$\pm$2.90&22.56$\pm$2.66&22.29$\pm$3.00&21.52$\pm$2.96&21.63\\%
\hline%
\multirow{2}{*}{20.0}&sup only&23.79$\pm$3.02&25.35$\pm$3.24&20.48$\pm$2.71&19.87$\pm$2.84&22.82$\pm$3.06&24.32$\pm$2.60&23.22$\pm$2.93&23.67$\pm$2.79&22.94\\%
&WarpDDF+RegCut&24.10$\pm$3.02&24.35$\pm$3.11&19.69$\pm$2.63&19.60$\pm$2.91&21.89$\pm$3.11&23.96$\pm$2.64&22.34$\pm$2.99&23.23$\pm$2.76&22.40\\%
\hline%
\multirow{2}{*}{50.0}&sup only&23.76$\pm$3.12&25.32$\pm$3.22&20.86$\pm$2.70&19.81$\pm$2.79&21.53$\pm$2.99&22.85$\pm$2.74&22.49$\pm$2.95&22.47$\pm$2.72&22.39\\%
&WarpDDF+RegCut&23.36$\pm$3.09&25.02$\pm$3.23&20.31$\pm$2.65&19.40$\pm$2.82&21.43$\pm$2.99&23.91$\pm$2.86&22.75$\pm$3.01&21.71$\pm$2.83&22.24\\%
\hline%
\multirow{1}{*}{100.0}&sup only&23.41$\pm$3.07&25.30$\pm$3.25&20.21$\pm$2.65&19.21$\pm$2.78&21.36$\pm$3.02&23.50$\pm$2.85&22.00$\pm$2.93&21.37$\pm$2.78&22.05\\%
\hline%
\end{tabular}%
    \label{tab:hd_by_class}
    \caption{95\% Hausdorff Distance (mm) of the proposed algorithms and its variants at when labels are available for 10\%, 20\%, 50\% and all training images.}
\end{table*}

\subsection{Dataset and Implementation Details}
The evaluation is performed on a male lower-pelvic MRI dataset~\cite{li2023prototypical} which includes 589 T2-weighted images acquired from the same number of patients. For each image, eight anatomical structures of planning interest, including bladder, bone, central gland, neurovascular bundle, obturator internus, rectum, seminal vesicle, peripheral zone were labelled. The images are randomly sampled into training and testing subsets in a 3:1 ratio, resulting in a training set with 442 images and a test set with 147 images. 

%\subsection{Implementation Details}
All images were normalised, resampled and centre-cropped to an image size of $256\times256\times48$, with a voxel dimension of $0.75\times0.75\times2.5$ during pre-processing. LocalNet~\cite{hu2018weakly} was adopted as the architecture for both student and teacher model with implementation from the MONAI repository~\cite{cardoso2022monai}. Image perturbation augmented fixed image through random rotation between -5 to 5 degree, scaling between 0.75 to 1.25 and translation between -20 to 20 voxels.
The models were trained for 4000 epochs with the first 2000 epochs warmed up on the labelled dataset using only the weakly-supervised loss before adding the weight and image consistency loss for semi-weakly supervision.
Training adopts an Adam optimiser starting at a learning rate of 0.00005 with a minibatch size of 1. The implementation code has been released at \codeurl.

Experiments are carried out at various labelled-unlabelled ratios $r$, such that $442 \times r$ images are provided with labels and $442 * (1-r)$ images are unlabelled. To investigate the effect of utilising unlabelled images, both results with only labelled images and with all training images are reported.
Classical intensity-based registration algorithms, `NiftyReg'~\cite{modat2014global}, are also compared as a non-learning baseline method.

\subsection{Ablation Studies on Registration Performance}

 To investigate the effectiveness of the proposed image perturbation methods, we report Dice and the 95$^{th}$ percentile Hausdorff distance (HD$^{95}$) results on all eight available anatomical ROIs, after the following variants of registration training strategies, 
\begin{itemize}
    \item 'NoAug': No image perturbation is applied such that teacher and student receive the same image pair.
    \item `WarpDDF': Only WarpDDF (Sec.~\ref{sec:method_image_warpddf}) is applied to the unlabelled fixed image.
    \item `RegCut': Only RegCut (Sec.~\ref{sec:method_image_regcut}) is used to generate the augmented unlabelled moving image.
    \item `WarpDDF+RegCut': Both WarpDDF and RegCut are applied (Sec.~\ref{sec:method_image_combined}).
    %\item `IntensitySimilarity': For comparison purpose, an additional baseline using only intensity-based unsupervised loss (here, SSD) on unlabelled image pairs~\cite{de2017end}.
\end{itemize}

\subsection{Computational Atlas for Pelvic MR Images}\label{sec:exp_atlas}

To demonstrate the potential clinical relevance of the semi-weakly supervised registration networks, a lower-pelvic atlas is constructed by registering clinical image samples, from the test set, using the different registration networks.
For a set of samples $\{I^i\}_i^N$, an atlas is initialised by choosing the sample that is the most similar to the average of all samples, based on the binary Dice score on their segmentation masks.
The final atlas is updated iteratively as the average of samples after registering each sample to the initialised atlas, until convergence is reached, as illustrated in Algorithm.~\ref{alg:atlas}.
In this experiment, we define a population diversity $\sigma^2_{pop}$ based on the computational atlas, as an example measure that can be used for subsequent clinical quantitative analysis. The adopted population diversity $\sigma^2_{pop}$ indicate the variance of the established correspondence, represented by the registration-produced displacement field, over all available subject samples.
It is calculated as the variance of paired distance between voxels registered to the same atlas locations, averaged over spatial voxel locations, $\sigma^2_{pop}=\mathbb{E}_{x,y,z}[\text{Var}(\{\|[u^i_{(1)xyz},u^i_{(2)xyz},u^i_{(3)xyz}]^{\top}-[u^j_{(1)xyz},u^j_{(2)xyz},u^j_{(3)xyz}]^{\top}\|^2\}_{i<j})]$
With $u_{dxyz}$ denotes the registration-produced displacement.
This measure is an example indicator of morphological and imaging features, which can potentially be predicting or confounding factors for clinical conditions. For example, the benign prostate enlargement is positively indicated by the size of the prostate gland, while the prostate cancer is negatively correlated with gland size. Statistics about gland volume have been found reliable to measure and straightforward to apply for positioning individual subjects in different (e.g. healthy and pathological) patient cohort populations, in other areas such as preclinical and human neuroimaging studies. The construction of the computational atlas should facilitate such population studies in pelvic organs. In this study, we present experimental results of the above-defined population diversity to quantify their differences due to semi-weakly-supervised registration, where specific hypothesis generation is made and discussed in Sec.~\ref{sec:result}, when interesting results are observed.

\begin{algorithm}\label{alg:atlas}
\caption{Atlas generation}
\SetKwInOut{Input}{input}\SetKwInOut{Output}{output}
\Input{Neural network $g$ with parameters $\theta$.\\
A set of sample $\{I^i\}_i^N$.\\
}
\Output{Atlas $A$.}
\For{$i \in {1,...,N}$}{
        Compute similarity between the sample and the average: $S^i=\text{Sim}(I^i, I^{avg})$
    }
Choose the sample with highest similarity: $c=\argmax_{i} \text{Sim}(I^i, I^{avg})$\\
Initialise atlas: $A_0=I^c$\\

\For{$i \in {1,...,N}$}{
    Predict DDF that register the sample to the current atlas: $u_t^i=g(I^i, A_0;\theta)$\\
    Warp the sample with the predicted DDF: $I^i_t=T(I^i, u_t^i)$\\
}
Update atlas: $A=\frac{1}{N}\sum_{i=0}^{N} I^i$
\end{algorithm}

\section{Results}\label{sec:result}

Table~\ref{tab:Dice_by_class} and \ref{tab:hd_by_class} respectively summarised Dice score and 95\% Hausdorff distance achieved by the proposed algorithms and its variants at when labels are available for 10\%, 20\%, 50\% and all training images. The proposed method with both strong augmentations achieved an improvement of 6.63\%, 2.67\% and 0.66\% in Dice score with p-value being 8.91e-37, 7.43e-12 and 2.81e-3 as well as a reduction of 1.37 mm, 0.61 mm and 0.24 mm in 95\% Hausdorff distance when labels are available for 10\%, 20\%, 50\% of the training images. Intuitively the improvement decreases as the labelled ratio increases. Notably, the proposed method trained with a labelled ratio 10\% outperformed the fully supervised method trained on the same amount of labelled images, at a labelled ratio of 50\%, and has a difference of only 2.28\% in Dice score to the fully supervised baseline.

Trained on a training set with 10\% of the images labelled, the semi-supervision without image perturbation with the proposed augmentations achieved an improvement of 2.78\% in Dice score. Adding the warp augmentation, RegCut augmentation and both augmentations respectively achieved a further improvement of 2.76\%, 2.16\% and 2.85\% with p-values being 1.93e-11, 9.47e-9 and 2.37e-10 (paired t-tests at a significance level of $\alpha$=0.05), respectively, proving the efficacy of both proposed augmentations.

\begin{figure}
    \centering
    \subfigure[]{
    \includegraphics[width=0.5\linewidth]{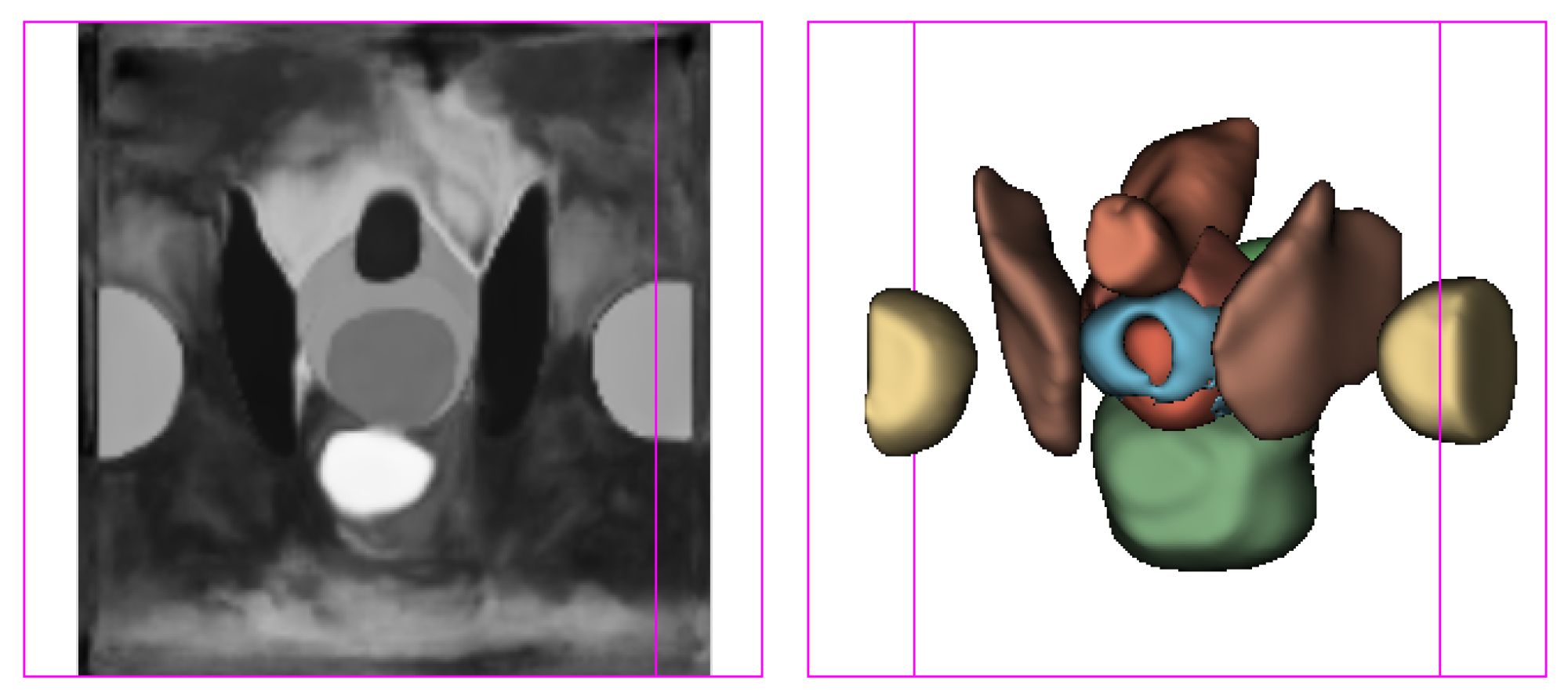}
    }
    \subfigure[]{
    \includegraphics[width=\linewidth]{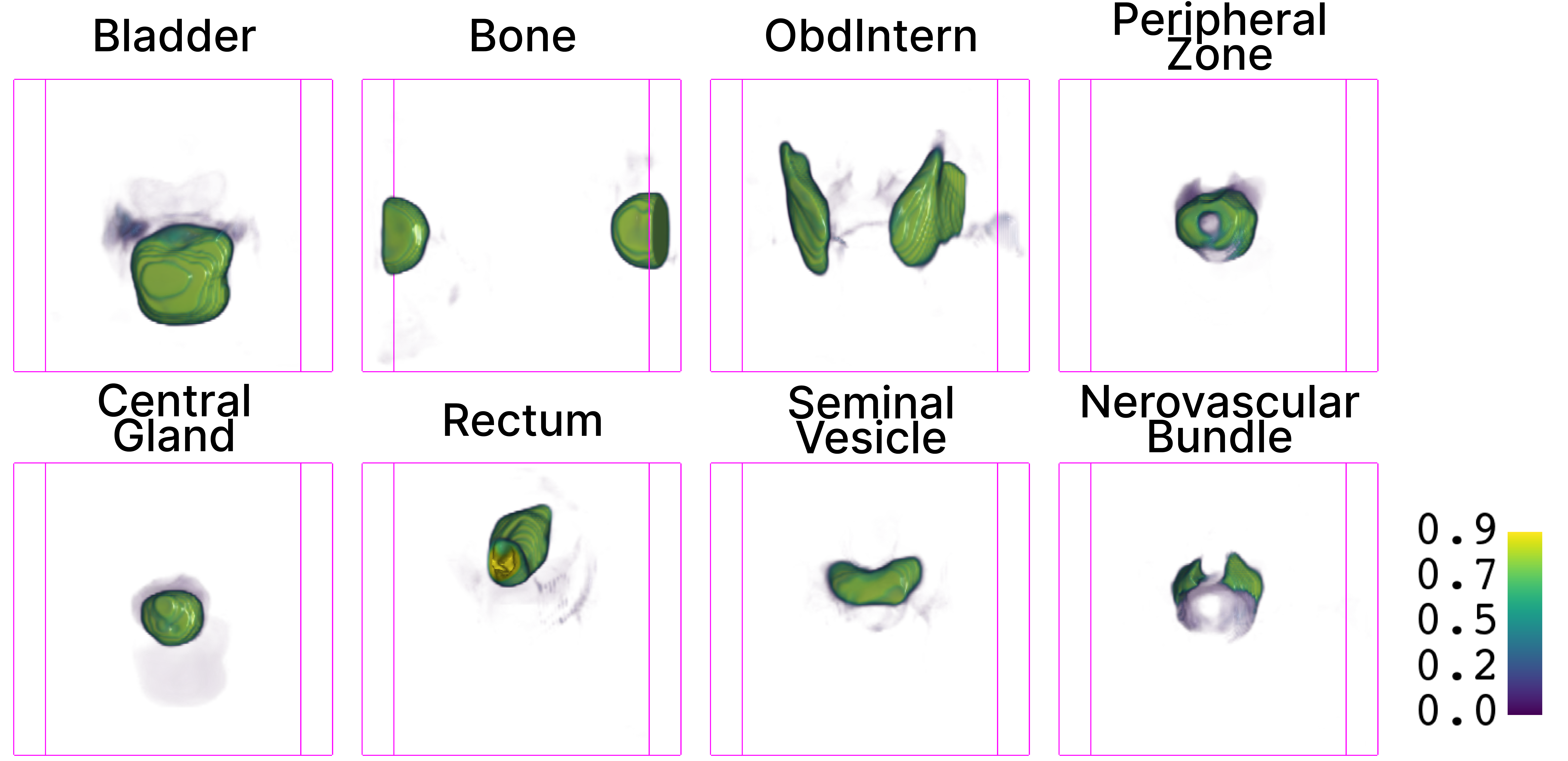}
    }
    \caption{\label{fig:atlas}(a) The resulting atlas and associated segmentation masks. (b) The probability maska of the atlas for the eight structures.}
    \label{fig:semi_weakly}
\end{figure}

\newcommand{\varbypop}{\begin{tabular}{|c|c|ccc|}%
\hline%
labelled&method&\multicolumn{3}{c|}{CG}\\%
ratio (\%)&&top 20\%&bottom 20\%&ratio\\%
\hline%
\multirow{5}{*}{10.0}&sup only&22.67&37.52&0.60\\%
&NoAug&22.94&40.59&0.57\\%
&WarpDDF&20.78&28.65&0.73\\%
&RegCut&30.05&49.45&0.61\\%
&WarpDDF+RegCut&17.68&23.93&0.74\\%
\hline%
\multirow{2}{*}{20.0}&sup only&25.76&41.42&0.62\\%
&WarpDDF+RegCut&20.85&33.55&0.62\\%
\hline%
\multirow{2}{*}{50.0}&sup only&25.03&35.88&0.70\\%
&WarpDDF+RegCut&21.86&31.00&0.71\\%
\hline%
\multirow{1}{*}{100.0}&sup only&19.06&24.00&0.79\\%
\hline%
\end{tabular}%
}
\begin{table}[t]
    \centering
    \begin{tabular}{|c|c|ccc|}%
\hline%
labelled&method&\multicolumn{3}{c|}{CG}\\%
ratio (\%)&&top 20\%&bottom 20\%&ratio\\%
\hline%
\multirow{5}{*}{10.0}&sup only&22.67&37.52&0.60\\%
&NoAug&22.94&40.59&0.57\\%
&WarpDDF&20.78&28.65&0.73\\%
&RegCut&30.05&49.45&0.61\\%
&WarpDDF+RegCut&17.68&23.93&0.74\\%
\hline%
\multirow{2}{*}{20.0}&sup only&25.76&41.42&0.62\\%
&WarpDDF+RegCut&20.85&33.55&0.62\\%
\hline%
\multirow{2}{*}{50.0}&sup only&25.03&35.88&0.70\\%
&WarpDDF+RegCut&21.86&31.00&0.71\\%
\hline%
\multirow{1}{*}{100.0}&sup only&19.06&24.00&0.79\\%
\hline%
\end{tabular}%
    \label{tab:population_variance}
    \caption{Population diversity across the top and bottom 20\% cohorts measured by central gland volume of the proposed algorithms and its variants at when labels are available for 10\%, 20\%, 50\% and all training images. The ratio between top and bottom 20\% population diversity is reported in the last column.}
\end{table}

Figure~\ref{fig:atlas} visualises the obtained atlas as described in Sec.~\ref{sec:exp_atlas}. 
Tables~\ref{tab:population_variance} clusters the computed population diversity based on their gland zonal volumes (a negative predictor of clinically significant cancer). Use the population diversity difference at the bottom row as a reference, between the top and bottom 20\% of the cases (bottom row), when supervised registration trained on 100\% labelled data. This difference, also indicated by the ratio of 0.79, decreases using fewer labelled data. Compared to alternative methods using a small amount (10\%) of labelled data, this difference was indeed increased by the combined DDF+RegCut, reflecting the proposed approaches' ability to recover the difference due to reduction in labelled data. 

\section{Discussion and conclusion}
This work proposed semi-weakly supervised methodologies with novel weight and image perturbation to reduce annotation required for segmentation-supervised registration. 
Experiments demonstrated the efficacy of using unlabelled data with the proposed approaches and augmentations on inter-subject multi-structural lower-pelvic MR registration. 
We further demonstrated that, using the proposed semi-weakly-supervised registration algorithms, a lower-pelvic anatomical atlas can be constructed and potentially benefits from the added unlabelled data. 

\section*{Acknowledgments}
This work was supported by the International Alliance for Cancer Early Detection, a partnership between Cancer Research UK [C28070/A30912; C73666/A31378], Canary Center at Stanford University, the University of Cambridge, OHSU Knight Cancer Institute, University College London and the University of Manchester. This work was also supported by 
the Wellcome/EPSRC Centre for Interventional and Surgical Sciences [203145Z/16/Z], and EPSRC [EP/T029404/1, EP/S021930/1].

\printbibliography

\section{Appendix - DDF composing and decomposing}\label{sec:app_ddf_compose}

\paragraph*{\textbf{Sequential DDFs}}
An output image $I^\text{(2)}$ is resampled by warping an original input image $I^\text{(0)}$, with two consecutive spatial transformations, represented by two DDFs, $U^\text{A}$ followed by $U^\text{B}$. 
\begin{equation}\label{eq:ddf_compose_i012}
    I^\text{(2)} = I^\text{(0)} \circ U^\text{A} \circ U^\text{B}
\end{equation}
Let $I^\text{{1}}$ denote the intermediate image, thus:
\begin{equation}\label{eq:ddf_compose_i01}
    I^\text{(1)} = I^\text{(0)} \circ U^\text{A}
\end{equation}
\begin{equation}\label{eq:ddf_compose_i12}
    I^\text{(2)} = I^\text{(1)} \circ U^\text{B}
\end{equation}
For the purpose of this proof, images $I^\text{(0)}(x_0,y_0,z_0)$, $I^\text{(1)}(x_1,y_1,z_1)$ and $I^\text{(2)}(x_2,y_2,z_2)$ are functions of spatial coordinates, in their respective image coordinate systems. 

Eq.~\ref{eq:ddf_compose_i01} represents the interpolation-based image resampling between the two imaging coordinates: 
\begin{equation}\label{eq:ddf_compose_x01}
    \begin{cases}
    x_1+u^A_{(d=1)}(x_1,y_1,z_1)=x_0\\
    y_1+u^A_{(d=2)}(x_1,y_1,z_1)=y_0\\
    z_1+u^A_{(d=3)}(x_1,y_1,z_1)=z_0\\
    \end{cases}
\end{equation}
where $u^A_{(d=1)}$, $u^A_{(d=2)}$ and $u^A_{(d=3)}$ are the DDFs in respective $x$-, $y$- and $z$-directions. In practice, they are functions of the spatial coordinates in the (transformed) image $I^\text{(1)}$ coordinate system (for querying $(x_0,y_0,z_0)$ locations in grid data interpolation). Analogously, Eq.~\ref{eq:ddf_compose_i12} leads to:
\begin{equation}\label{eq:ddf_compose_x12}
    \begin{cases}
    x_2+u^B_{(d=1)}(x_2,y_2,z_2)=x_1\\
    y_2+u^B_{(d=2)}(x_2,y_2,z_2)=y_1\\
    z_2+u^B_{(d=3)}(x_2,y_2,z_2)=z_1\\
    \end{cases}
\end{equation}
Therefore, the sequential transformation, applying $U^\text{A}$ and $U^\text{B}$ as in Eq.~\ref{eq:ddf_compose_i012}, obtains:
\begin{equation}\label{eq:ddf_sequence_final}
    \begin{split}
        x_0=x_2 &+u^B_{(d=1)}(x_2,y_2,z_2)+u^A_{(d=1)}(x_1,y_1,z_1)\\
        =x_2 &+u^B_{(d=1)}(x_2,y_2,z_2)+u^A_{(d=1)}\\
        (&x_2+u^B_{(d=1)}(x_2,y_2,z_2),\\
        &y_2+u^B_{(d=2)}(x_2,y_2,z_2),\\
        &z_2+u^B_{(d=3)}(x_2,y_2,z_2))
    \end{split}
\end{equation}

\paragraph*{\textbf{Composed DDFs}}
Now we consider composing $U^\text{A}$ and $U^\text{B}$, by first defining a new composed spatial transformation $U^\text{C}$, such that:
\begin{equation}\label{eq:ddf_compose_i02}
    I^\text{(2)} = I^\text{(0)} \circ U^\text{C}
\end{equation}
Using the same notation as in Eqs.~\ref{eq:ddf_compose_x01} and~\ref{eq:ddf_compose_x12}, this spatial transformation in Eq.~\ref{eq:ddf_compose_i02} represents:
\begin{equation}
    \begin{cases}
    x_2+u^C_{(d=1)}(x_2,y_2,z_2)=x_0\\
    y_2+u^C_{(d=2)}(x_2,y_2,z_2)=y_0\\
    z_2+u^C_{(d=3)}(x_2,y_2,z_2)=z_0\\
    \end{cases}
\end{equation}
Let $U^\text{C} = U^\text{B} + U^\text{D}$, where $U^\text{D}=U^\text{A} \circ U^\text{B}$ is a generalisation of the resampling in Eq.~\ref{eq:ddf_compose_i12}, applying $U^\text{B}$ on individual displacement components of $U^\text{A}$, i.e. $d=1$, $d=2$ and $d=3$: 
\begin{equation}
    \begin{cases}
    x_D+u^B_{(d=1)}(x_D,y_D,z_D)=x_A\\
    y_D+u^B_{(d=2)}(x_D,y_D,z_D)=y_A\\
    z_D+u^B_{(d=3)}(x_D,y_D,z_D)=z_A\\
    \end{cases}
\end{equation}
where $(x_A,y_A,z_A)$ and $(x_D,y_D,z_D)$ are spatial coordinates in the $U^\text{A}$ and $U^\text{D}$ DDF coordinate systems, respectively. 

For resampling $I^\text{(2)}$, the image coordinate system $I^\text{(2)}$ defines the $U^\text{D}$ coordinate system, that is:
\begin{equation}
    \begin{cases}
    x_2+u^B_{(d=1)}(x_2,y_2,z_2)=x_A\\
    y_2+u^B_{(d=2)}(x_2,y_2,z_2)=y_A\\
    z_2+u^B_{(d=3)}(x_2,y_2,z_2)=z_A\\
    \end{cases}
\end{equation}
Therefore, the composed transformation, applying $U^\text{C}$ as in Eq.~\ref{eq:ddf_compose_i02}, obtains:
\begin{equation} \label{eq:ddf_compose_final}
    \begin{split}
        x_0=x_2 &+u^{C}_{(d=1)}(x_2,y_2,z_2)\\
        =x_2 &+u^B_{(d=1)}(x_2,y_2,z_2)+u^{D}_{(d=1)}(x_A,y_A,z_A)\\
        =x_2 &+u^B_{(d=1)}(x_2,y_2,z_2)+u^A_{(d=1)}\\
        (&x_2+u^B_{(d=1)}(x_2,y_2,z_2),\\
        &y_2+u^B_{(d=2)}(x_2,y_2,z_2),\\
        &z_2+u^B_{(d=3)}(x_2,y_2,z_2))
    \end{split}
\end{equation}

\paragraph*{\textbf{Decomposed DDFs}}
Finally, comparing Eq.~\ref{eq:ddf_compose_final} and Eq.~\ref{eq:ddf_sequence_final}, which also holds for $y_0$ and $z_0$, shows that:
\begin{equation}\label{eq:ddf_compose}
    I^\text{(2)} = I^\text{(0)} \circ (U^\text{B} + U^\text{A} \circ U^\text{B}) = I^\text{(0)} \circ U^\text{A} \circ U^\text{B}
\end{equation}

\iffalse
TODO: $I$ could be DDF

\begin{equation}
    I^{out} = T(T(I^{in},u^1),u^2)
\end{equation}
denote $I^{mid}=T(I^{in},u^1)$, at any position $(x,y,z)$:
\begin{equation}
    I^{out}_{(x,y,z)}=I^{mid}_{(x_1,y_1,z_1)}=I^{in}_{(x_2,y_2,z_2)}
\end{equation}
where
\begin{equation}
    \begin{cases}
    x_1=x+u^2_{(0,x,y,z)}\\
    y_1=y+u^2_{(1,x,y,z)}\\
    z_1=z+u^2_{(2,x,y,z)}\\
    \end{cases}
\end{equation}
and
\begin{equation}
    \begin{cases}
    x_2=x_1+u^1_{(0,x_1,y_1,z_1)}\\
    y_2=y_1+u^1_{(1,x_1,y_1,z_1)}\\
    z_2=z_1+u^1_{(2,x_1,y_1,z_1)}
    \end{cases}
\end{equation}
It could be derived that
\begin{equation}
    \begin{split}
        x_2&=x+u^2_{(0,x,y,z)}+u^1_{(0,x_1,y_1,z_1)}\\
        &=x+u^2_{(0,x,y,z)}+
        u^1_{(0,
        x+u^2_{(0,x,y,z)},
        y+u^2_{(1,x,y,z)},
        z+u^2_{(2,x,y,z)})}\\
        &=x+u^2_{(0,x,y,z)}+T(u^1,u^2)_{(0,x,y,z)}\\
        &=x+u^{total}_{(0,x,y,z)} \text{, where }u^{total}=u^2+T(u^1,u^2)
    \end{split}
\end{equation}
Similarly
\begin{equation}
    \begin{cases}
    y_2=y+u^{total}_{(1,x,y,z)}\\
    z_2=z+u^{total}_{(2,x,y,z)}
    \end{cases}
\end{equation}
Therefore 
\begin{equation}
    I^{out}=T(I^{in},u^{total})
\end{equation}

\fi
\end{document}